\documentclass[preprint2,twocolumn,times,tighten]{aastex63}

\usepackage{graphicx,times}
\usepackage{subfigure}
\usepackage{threeparttable}
\usepackage{booktabs}
\usepackage{CJK}

\usepackage{amsmath}
\usepackage{cases}
\usepackage{longtable}
\usepackage{hyperref}
\usepackage{epstopdf}
\usepackage{amsmath,bm}
\usepackage{amssymb}
\usepackage{natbib}
\usepackage{morefloats}
\usepackage{multirow}
\usepackage{array}
\usepackage{verbatim}
\usepackage{color}
\usepackage{lineno}

\begin{document}
\begin{CJK*}{UTF8}{gbsn}

\title{On the injection scale of the turbulence in the partially ionized very local interstellar medium}

\author[0000-0002-0458-7828]{Siyao Xu (徐思遥)}
\affiliation{Institute for Advanced Study, 1 Einstein Drive, Princeton, NJ 08540, USA; sxu@ias.edu
\footnote{NASA Hubble Fellow}}

\author[0000-0003-3556-6568]{Hui Li (李晖)}
\affiliation{Los Alamos National Laboratory, NM 87545, USA; hli@lanl.gov}

\begin{abstract}

The cascade of magnetohydrodynamic
(MHD) turbulence is subject to ion-neutral collisional damping and neutral viscous damping in the partially ionized interstellar medium. 
By examining the damping effects in the warm and partially ionized local interstellar medium,
we find that the interstellar turbulence is damped by neutral viscosity 
{ at $\sim 261$ au}
and cannot account for the turbulent magnetic fluctuations detected by {\it Voyager 1} and {\it 2}.
The MHD turbulence { measured by {\it Voyager} in the very local interstellar medium (VLISM)} should be locally injected in the regime where ions are decoupled from neutrals for its cascade to survive the damping effects. With the imposed ion-neutral decoupling condition
and the strong turbulence condition for the observed Kolmogorov magnetic energy spectrum, 
we find that the turbulence in the VLISM is sub-Alfv\'{e}nic,
{and its largest possible injection scale is $\sim 194$ au. }

\end{abstract}

\section{Introduction}

Turbulent magnetic fluctuations following a Kolmogorov spectrum were observed by {\it Voyager 1} and {\it 2}
in the outer heliosheath 
{ \citep{Burla18, Zhao20,Lee20,Frat21,Bur22}}.
While the source(s) of the turbulence in the very local interstellar medium 
(VLISM) are under debate 
\citep{Holz89,Zank15, Zank19},
it is believed to play a crucial role in
affecting the transport of 
energetic particles and cosmic rays
\citep{LaO09,Stone13,Krim13,Lop17,Frate22}, 
and the structure of the Interstellar Boundary Explorer (IBEX) ribbon
\citep{GiaJo15,Zirn20}. 
In particular, modeling by \cite{Zirn20} suggests
that the turbulent magnetic fields with scales $< 100$ au are
important for producing the ribbon structure similar to {\it IBEX} observations,
and such turbulence is likely not of pristine interstellar origin,
whereas turbulent fluctuations at scales $\geq 100$ au produce 
features inconsistent with {\it IBEX} observations. 

Turbulence and turbulent magnetic fields are ubiquitous in the ISM
(e.g., \citealt{Armstrong95,CheL10,Gae11,XuZ17,Laz18}), 
with the turbulent energy injected by supernova explosions 
on large length scales ($\sim 100$ pc;
\citealt{Brei17}) 
and cascading down toward smaller and smaller scales. 
In the warm and partially ionized local interstellar medium (LISM) 
\citep{Fris11,Sla08},
the magnetohydrodynamic (MHD) turbulence is subject 
to the damping effects due to the frictional collisions between 
ions and neutrals (i.e., ion-neutral collisional damping) and 
among neutrals (i.e. neutral viscosity)
\citep{XLp17}. 
The damping effects cause the cutoff of MHD turbulence 
cascade when the damping rate exceeds the cascading rate of MHD turbulence.

The linear analysis of MHD waves in a partially ionized medium was performed by 
\citet{Kulsrud_Pearce} and 
more recently by e.g., 
\citet{Pudr90,Bals96,Mou11,Zaqa11,Sol13}.
Different from linear MHD waves, 
MHD turbulence is characterized by the nonlinear cascade of turbulent energy, with scale-dependent turbulence anisotropy and limited timescale of turbulent motions
\citep{GS95,LV99}. 
The ion-neutral collisional (IN) damping of MHD turbulence in a highly ionized medium at a high plasma $\beta$ was studied by 
\citet{LG01}.
The neutral viscous (NV) damping of MHD turbulence was analyzed by 
\citet{LVC04}, 
which tends to dominate over the IN damping 
toward a higher ionization fraction and a higher temperature
\citep{XLp17}.
\citet{XLY14,Xuc16}
performed a general analysis including both damping effects of MHD turbulence in different interstellar phases with varying ionization fractions.

The range of length scales for the existence of MHD turbulence depends on the coupling state between ions and neutrals and the corresponding damping effects. Irrespective of the origin, the MHD turbulence measured in the VLISM should survive the damping in the partially ionized medium. In this work, we will examine the damping of the interstellar MHD turbulence driven at large scales.
More importantly, we will 
explore the constraint imposed by the damping effects on the locally driven MHD (LMHD) turbulence in the VLISM,
as observed by {\it Voyager}. 
In Section 2, we analyze 
the damping of the interstellar turbulence driven in the strongly coupled regime 
and the injection of the LMHD turbulence in the decoupled regime with weak damping. 
In Section 3, 
we determine the turbulence regime and the 
largest injection scale of the LMHD turbulence
constrained by the ion-neutral decoupling condition. 
Discussion and conclusions are presented in Sections 4 and 5.


\section{ Source for turbulence in the VLISM}

{ There are three possible sources for the turbulent magnetic fluctuations measured by {\it Voyager}, including the interstellar turbulence driven by supernova explosions at $\sim 100$ pc, the turbulence driven in the Local Interstellar Cloud at $\sim 2$ pc
\citep{Zank19}, 
and the LMHD turbulence in the VLISM. 
We will first focus on the 
former two possible sources, for which the turbulence is driven on large scales in the regime where neutrals and ions are strongly coupled together (Section \ref{ssec: nvin}). 
For the local source, the turbulence with a small energy injection scale is expected to arise in the regime where ions are decoupled from neutrals (Section \ref{ssec:westt}).  
}

\subsection{Damping of interstellar turbulence driven in strongly coupled regime }
\label{ssec: nvin}
The interstellar turbulence is driven by supernova explosions at $L \sim 100$ pc
\citep{Brei17}
and cascades down to smaller and smaller scales
\citep{Chep10,Yuen22}.
{ Its damping scale varies in different interstellar phases and regions, depending on the local physical conditions. }
Given the approximate energy equipartition between turbulence and magnetic fields in the ISM
expected from small-scale nonlinear turbulent dynamo 
\citep{XL16}
and indicated by observational measurements 
\citep{Pat22},
we consider that the interstellar turbulence is trans-Alfv\'{e}nic, i.e., 
$M_A = V_L/V_A \approx 1$.
Here $M_A$ is the { turbulent} Alfv\'{e}n Mach number, $V_L$ is the turbulent velocity at the injection scale $L$ of turbulence, 
$V_A = B/\sqrt{4\pi\rho}$ is the Alfv\'{e}n speed, 
$B$ is the magnetic field strength, 
$\rho = \rho_i + \rho_n$ is the total mass density,
$\rho_i = m_H n_i = m_H n_e$ is the mass density of ions, 
$m_H$ is the hydrogen atomic mass, $n_i$ and $n_e$ are the number densities of ions and electrons,
$\rho_n = m_H n_H$ is the mass density of neutrals, and $n_H$ is the number density of neutrals. 

In a partially ionized medium, 
MHD turbulence is subject to both IN damping and NV damping
\citep{LVC04,XLp17}.
{ They both} depend on the coupling state between ions and neutrals.
{ When the driving rate of turbulence is lower than the neutral-ion collisional frequency $\nu_{ni} = \gamma_d \rho_i$, 
where $\gamma_d = 5.5 \times10^{14}$ cm$^3$ g$^{-1}$ s$^{-1}$ is the drag coefficient 
\citep{Drai83,Shu92},
the MHD turbulence is driven in strongly coupled ions and neutrals, i.e., the strongly coupled regime.}
Neutrals decouple from ions when the cascading rate of MHD turbulence $\tau_\text{cas}^{-1}$ becomes 
larger than $\nu_{ni}$.
The neutral-ion decoupling entails significant collisional friction, causing the damping of MHD turbulence and the cutoff of 
its energy cascade. 
The corresponding damping scale for Alfv\'{e}nic turbulence is
(\citealt{XLY14,Xuc16}; see Appendix \ref{app:dam} for the derivation)
\begin{equation}\label{eq: kdam}
    l_{\text{dam,IN},\perp} = \Big(\frac{2\nu_{ni}}{\xi_n}\Big)^{-\frac{3}{2}} L^{-\frac{1}{2}} V_L^{\frac{3}{2}},                             
\end{equation}
where $\xi_n = \rho_n/\rho$.
The subscript $\perp$ means that the length scale is measured perpendicular to the local magnetic field. 
As the energy cascade of MHD turbulence is mostly in the perpendicular direction, 
$l_{\text{dam,IN},\perp}$ corresponds to the cutoff scale of MHD turbulence. 

In the case when the NV damping dominates over the IN damping, the damping of MHD turbulence occurs in the strongly coupled regime, with the damping scale  
(\citealt{XLp17}; see Appendix \ref{app:dam})
\begin{equation}
l_{\text{dam,NV},\perp}=\Big(\frac{\xi_n}{2}\Big)^\frac{3}{4}\nu_n^\frac{3}{4} L^\frac{1}{4} V_L^{-\frac{3}{4}},
\end{equation}
where $\nu_n = v_\text{th}/(n_H \sigma_{nn})$ is the kinematic viscosity in neutrals, $v_\text{th}$ is the neutral thermal speed, and $\sigma_{nn}$ is the collisional cross-section of neutrals. 
We note that $l_{\text{dam,NV},\perp}$ is in fact the damping scale of the turbulent kinetic energy spectrum. 
The magnetic fluctuations in the sub-viscous range below $l_{\text{dam,NV},\perp}$ is termed new regime of MHD turbulence 
\citep{LVC04}
(see Section \ref{ssec: newmhd}).

We now discuss whether the magnetic turbulence in the VLISM can come from the interstellar turbulence
that is injected by supernova explosions.
The LISM near the sun is warm, low-density, and partially ionized, with the temperature 
$T \approx 6300 $ K,
$n_H \approx 0.2 $ cm$^{-3}$, and $n_e \approx 0.07$ cm$^{-3}$
\citep{Sla08,Swa20}.  
In addition, we adopt $L_\text{ISM} \approx 100$ pc, $V_{L,\text{ISM}} \approx V_A$, $B \approx 5~\mu$G 
as the typical driving conditions of interstellar turbulence and interstellar magnetic field strength 
\citep{Crut10}.
With these high temperature and moderate ionization fraction,
we find that 
\begin{equation}\label{eq: ismdampe}
\begin{aligned}
   l_{\text{dam,IN},\perp} \approx  &7.6\times10^{13}~\text{cm} \Big(\frac{n_e}{0.07~\text{cm}^{-3}}\Big)^{-\frac{3}{2}}
   \Big(\frac{n_H}{0.2~\text{cm}^{-3}}\Big)^{-\frac{3}{4}} \\
    &\Big(\frac{L_\text{ISM}}{100~\text{pc}}\Big)^{-\frac{1}{2}} \Big(\frac{B}{5~\mu \text{G}}\Big)^{\frac{3}{2}}
\end{aligned}
\end{equation}
is much smaller than 
\begin{equation}\label{eq:ismdamnv}
\begin{aligned}
l_{\text{dam,NV},\perp}
&\approx 3.9\times10^{15}~\text{cm} \Big(\frac{T}{6300~\text{K}}\Big)^\frac{3}{8} \Big(\frac{n_H}{0.2~\text{cm}^{-3}}\Big)^{-\frac{3}{8}} \\ &~~~~\Big(\frac{L_\text{ISM}}{100~\text{pc}}\Big)^\frac{1}{4} \Big(\frac{B}{5~\mu \text{G}}\Big)^{-\frac{3}{4}},
\end{aligned}
\end{equation}
where we assume $n_H+n_e \sim n_H$, and $\sigma_{nn}\approx 10^{-14}$ cm$^2$
{ \citep{Krst98,VrKr13}} 
is adopted. 
{ We note that the gradients of quantities in the outer heliosheath 
\citep{Zank13}
cause uncertainties in the above estimates. 
By assuming the uncertainties $\sigma(n_e) \sim 0.01$~cm$^{-3}$, 
$\sigma(n_H) \sim 0.01$~cm$^{-3}$,
$\sigma(B) \sim 1~\mu$G, $\sigma(T) \sim 1000$~K, and $\sigma(L_\text{ISM})\sim 10~$pc,
we find $\sigma(l_{\text{dam,IN},\perp}) \sim 2.8\times10^{13}~$cm 
and $\sigma(l_{\text{dam,NV},\perp})\sim 6.4\times10^{14}~$ cm.}

{ For the interstellar turbulence driven in the Local Interstellar Cloud, the turbulent velocity of a few km s$^{-1}$ is similar to that from the cascade of supernova-driven turbulence at $\sim 2$ pc
\citep{Spang11}.
It might be a part of the global cascade of the interstellar turbulence. 
Therefore, it is also subject to the damping effects in the partially ionized LISM and damped at the damping scale similar to that of the supernova-driven interstellar turbulence.
}

{ The above calculations show that the damping of the MHD turbulence 
in the LISM 
from the interstellar origin
is dominated by NV in the strongly coupled regime. 
The corresponding damping scale is 
$\sim 3.9 \times 10^{15}$ cm, i.e., $261$ au.}
This means that the turbulence observed by {\it Vogayer 1} and {\it 2}
is unlikely due to the pristine interstellar origin.

\subsection{LMHD turbulence in decoupled regime}\label{ssec:westt}

As discussed above, for the MHD turbulence injected in the strongly coupled regime, its cascade is cut off either by the IN or NV damping. When ions are decoupled from neutrals, however,
the driven MHD turbulence is no longer subject to the NV damping, 
and the IN damping becomes constantly weak \citep{Xuc16}. 
The MHD turbulence injected in ions that are decoupled from neutrals, i.e., 
LMHD turbulence in the decoupled regime, 
is not cut off due to the damping effects arising in a partially ionized medium.

Based on the in-situ measurements of turbulent magnetic energy spectrum by {\it Voyager 1} in the VLISM
{ \citep{Lee20},
we have the ratio of the turbulent component to the average magnetic field strength 
$\delta B_\text{obs} / B \approx 0.06$
measured at $l_\text{obs} \approx 3\times10^{14}$ cm}
(corresponding to the smallest wavenumber of the measured spectrum and { assumed to be in the inertial range of turbulence}).
By assuming that the magnetic fluctuations are mainly induced by Alfv\'{e}nic turbulence
\citep{CL02_PRL,Hu22,Lee20}, 
the local turbulent velocity can be estimated as 
\begin{equation}\label{eq: vlocvog}
\begin{aligned}
     v_\text{obs}&= \frac{\delta B_\text{obs}}{B} V_{Ai} \\
          &= 2.5 ~\text{km s}^{-1} \Big(\frac{\delta B_\text{obs} /B}{0.06}\Big)  \Big(\frac{n_e}{0.07~\text{cm}^{-3}}\Big)^{-\frac{1}{2}}\Big(\frac{B}{5~\mu\text{G}}\Big).
\end{aligned}
\end{equation}
{ By assuming $\sigma(\delta B_\text{obs}) \sim 0.1~\mu$G and the uncertainties of other parameters (see Section \ref{ssec: nvin}), we find 
$\sigma(v_\text{obs}) \sim 0.85~$km s$^{-1}$.
The large uncertainties in our calculations are mainly caused by the large uncertainties in magnetic field strength measurements 
\citep{Burla18}.}
Such a large turbulence level at the measured length scale cannot be accounted for by the interstellar turbulence, which is cut off at a larger length scale $l_{\text{dam,NV},\perp}$ (Eq. \eqref{eq:ismdamnv}).
So the measured turbulence is likely to be driven in the VLISM. { Given $v_\text{obs} / l_\text{obs} (\approx 8.3\times10^{-10}~\text{s}^{-1}) > \nu_{in} (\approx 1.8\times10^{-10}~\text{s}^{-1})$,} the LMHD turbulence measured by {\it Voyager 1} should be injected in the decoupled regime.


\section{ Constraint on the injection scale of the LMHD turbulence in the VLISM}

\begin{figure*}[ht]
\centering
\subfigure[$M_A >1$, isotropic injection]{
   \includegraphics[width=5.2cm]{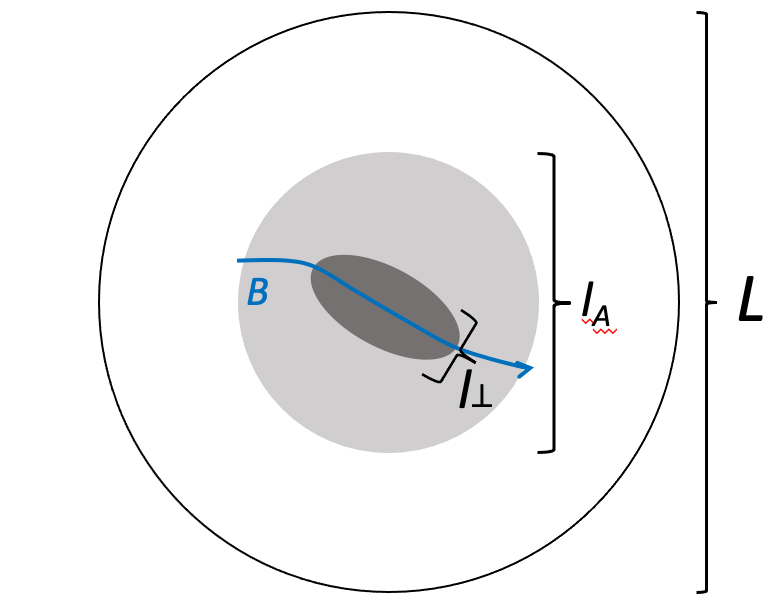}\label{fig: supA}}
\subfigure[$M_A<1$, isotropic injection]{
   \includegraphics[width=6cm]{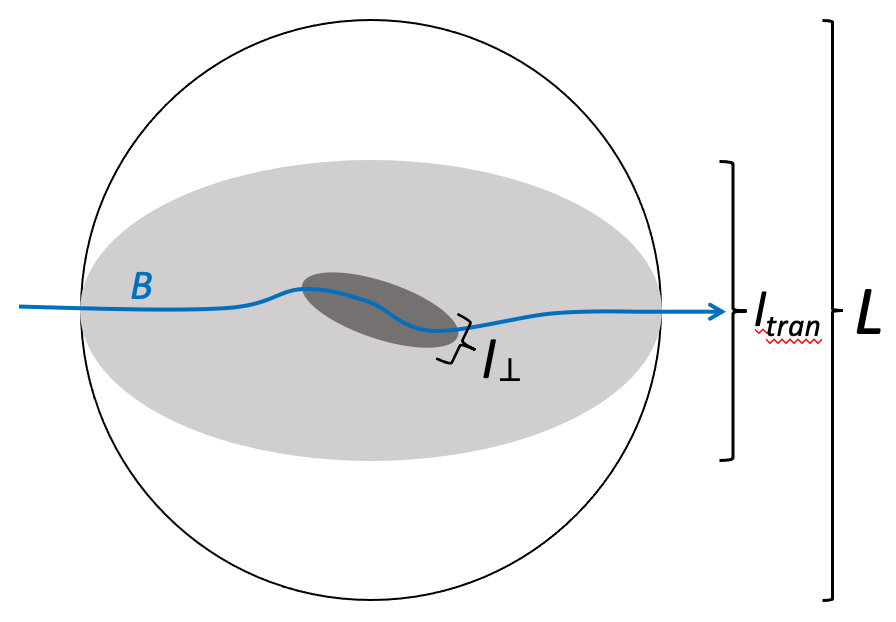}\label{fig: subAiso}}
\subfigure[$M_A<1$, anisotropic injection]{
   \includegraphics[width=5.7cm]{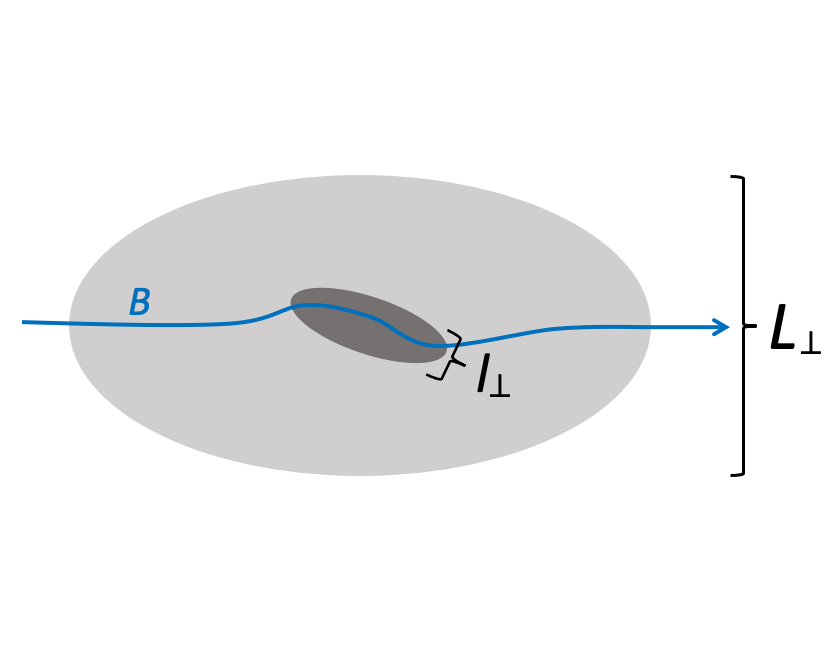}\label{fig: subAani}}
\caption{ Illustration for turbulent eddies in super-Alfv\'{e}nic (a) and sub-Alfv\'{e}nic (b,c) turbulence.
Strong MHD turbulence is indicated by shaded regions. 
With anisotropic injection scale of turbulence in (c), the entire turbulent cascade can be in the strong turbulence regime. 
The blue line indicates the magnetic field.  }
\label{fig: tur}
\end{figure*}

We now discuss the constraint on the injection scale $L$ of the LMHD turbulence.
{ The three cases we consider are (a) super-Alfv\'{e}nic turbulence with isotropic injection scale, (b) sub-Alfv\'{e}nic turbulence with isotropic injection scale, and (c) sub-Alfv\'{e}nic turbulence with anisotropic injection scale.
In Fig. \ref{fig: tur}, we illustrate the scalings of super- and sub-Alfv\'{e}nic turbulence
(see also Appendix \ref{app:ani})
for isotropic and anisotropic injection.}

The condition for the LMHD turbulence to arise in ions alone is 
\begin{equation}\label{eq: con}
    \Gamma_L > \nu_{in}, 
\end{equation}
where $\Gamma_L$ is the driving rate of the turbulence, i.e., the cascading rate of turbulence at $L$, 
and $\nu_{in} = \gamma_d \rho_n$ is the ion-neutral collisional frequency.

{\it Case} (a).
If the driven turbulence is super-Alfv\'{e}nic, i.e., $M_A = V_L /V_{Ai} >1$, 
where $V_{Ai} = B/\sqrt{4\pi\rho_i}$ is the Alfv\'{e}n speed in ions,
$V_L$ is related to $v_l$ by 
\citep{LV99}
\begin{equation}\label{eq: supasc}
   v_l = V_L \Big(\frac{l}{L}\Big)^\frac{1}{3}, 
\end{equation}
where $v_l$ is the local turbulent velocity measured at length scale $l (<L)$. 
With the cascade of turbulent energy, the turbulent velocity becomes equal to $V_{Ai}$
at the Alfv\'{e}nic scale $l_A = LM_A^{-3}$
\citep{Lazarian06}. Below $l_A$, the effect of magnetic fields on turbulence becomes important, resulting in turbulence anisotropy
(see Fig. \ref{fig: tur}). Therefore, 
$l$ in the above equation should be replaced by $l_\perp$ when $l$ is smaller than $l_A$, where $l_\perp$ is the length scale measured perpendicular to the local magnetic field
\citep{CV00}. 
By using $\Gamma_L = V_L / L$ and the scaling relation in Eq. \eqref{eq: supasc}, 
the condition Eq. \eqref{eq: con} in this case becomes 
\begin{equation}\label{eq: coninjsup}
\begin{aligned}
  & L < \Big(\nu_{in} v_l^{-1} l^\frac{1}{3}\Big)^{-\frac{3}{2}}, ~ l_A <l < L, \\
  & L < \Big(\nu_{in} v_l^{-1} l_\perp^\frac{1}{3}\Big)^{-\frac{3}{2}}, ~ l < l_A.
\end{aligned}
\end{equation}
The above expressions provide constraints on the maximum $L$ of the LMHD turbulence in the decoupled regime if it is super-Alfv\'{e}nic.

The in-situ measurements show that 
the LMHD turbulence has $v_l \approx v_\text{obs}$ at $l_\perp \approx l_\text{obs}$ (Section \ref{ssec:westt}). 
{ We consider the largest observed scale $l_\text{obs}$ as the perpendicular scale because the trajectory of {\it Voyager 1} 
is nearly perpendicular to the background magnetic field 
\citep{Izmo20,Frat21,Dia22}.}
$v_l$ is smaller than $V_{Ai}$ (see Eq. \eqref{eq: vlocvog}) and related to $V_{Ai}$ by 
\begin{equation}
    v_l = V_{Ai} \Big(\frac{l_\perp}{l_A}\Big)^\frac{1}{3}, 
\end{equation}
yielding 
\begin{equation}\label{eq: perla}
\begin{aligned}
   l_A &= l_\perp \Big(\frac{v_l}{V_{Ai}}\Big)^{-3}  \\
   &=  1.4\times10^{18} ~\text{cm} \Big(\frac{n_e}{0.07~\text{cm}^{-3}}\Big)^{-\frac{3}{2}}\Big(\frac{v_l}{2.5~\text{km s}^{-1}}\Big)^{-3}\\
   &~~~~  \Big(\frac{l_\perp}{3.0\times10^{14}~\text{cm}}\Big)
           \Big(\frac{B}{5~\mu\text{G}}\Big)^{3},
\end{aligned}
\end{equation}
{ with the estimated uncertainty $\sigma(l_A) \sim 1.7\times10^{18}~$cm.}
The condition in Eq. \eqref{eq: coninjsup} for super-Alfv\'{e}nic turbulence can be written explicitly as,
\begin{equation}
\begin{aligned}
  L < &2.9\times10^{15}~\text{cm} \Big(\frac{n_H}{0.2~\text{cm}^{-3}}\Big)^{-\frac{3}{2}} 
  \Big(\frac{v_l}{2.5~\text{km s}^{-1}}\Big)^\frac{3}{2} \\
  &\Big(\frac{l_\perp}{3.0\times10^{14}~\text{cm}}\Big)^{-\frac{1}{2}},
\end{aligned}
\end{equation}
{ with the estimated uncertainty $\sigma(L)\sim 1.5\times10^{15}~$cm.
The super-Alfv\'{e}nic condition $M_A>1$
requires $L >  l_A$}, which is not satisfied by the above
values. We conclude that the LMHD turbulence cannot be super-Alfv\'{e}nic.

{\it Case} (b). If the driven turbulence is sub-Alfv\'{e}nic, i.e., $M_A = V_L /V_{Ai} <1$, 
the MHD turbulence is weak 
{ with weak interactions between counterpropagating Alfv\'{e}n wave packets
\citep{Galt00}}
over scales $[L, l_\text{tran}]$, where 
$l_\text{tran} = LM_A^2$ is the perpendicular transition scale from weak to strong MHD turbulence
\citep{LV99}
(see Fig. \ref{fig: tur}).  
For weak turbulence, 
there is no parallel cascade.
Its cascade to smaller perpendicular scales strengthens until the cascade becomes strong
\citep{GS97}.
The scaling of weak turbulence follows $v_l = V_L (l_\perp/L)^{1/2}$
\citep{LV99}, while
in the strong turbulence regime, the local turbulent velocity follows the scaling 
\citep{LV99}
\begin{equation}\label{eq: subvsc}
    v_l =  V_L \Big(\frac{l_\perp}{L}\Big)^\frac{1}{3} M_A^\frac{1}{3}. 
\end{equation}
For the weak turbulence at $L$, there is 
\begin{equation}\label{eq: weinc}
   \Gamma_L = \frac{V_L}{L} M_A.
\end{equation}
By combining Eqs. \eqref{eq: subvsc} and \eqref{eq: weinc}, the condition in Eq. \eqref{eq: con} becomes 
\begin{equation}\label{eq: coninsub}
    L < \nu_{in}^{-2} v_l^3 l_\perp^{-1} V_{Ai}^{-1}. 
\end{equation}

{ The spectral indices of magnetic fluctuations in weak and strong MHD turbulence are $-2$ and $-5/3$, respectively.}
With the Kolmogorov slope ($-5/3$) found for the measured magnetic energy spectrum
{ \citep{Burla18,Lee20}}, we consider that the locally measured turbulence 
is in the strong MHD turbulence regime. 
For the driven sub-Alfv\'{e}nic turbulence, the condition in Eq. \eqref{eq: coninsub} gives 
\begin{equation}
\begin{aligned}
   L < &3.6\times10^{14}~\text{cm}  
   \Big(\frac{n_e}{0.07~\text{cm}^{-3}}\Big)^\frac{1}{2} \Big(\frac{n_H}{0.2~\text{cm}^{-3}}\Big)^{-2} \\
   &\Big(\frac{v_l}{2.5~\text{km s}^{-1}}\Big)^{3} \Big(\frac{l_\perp}{3.0\times10^{14}~\text{cm}}\Big)^{-1} \Big(\frac{B}{5~\mu\text{G}}\Big)^{-1},
\end{aligned}
\end{equation}
{ with $\sigma(L)\sim 3.8\times10^{14}~$cm.}
By using Eqs. \eqref{eq: subvsc} and \eqref{eq: weinc}, the condition in Eq. \eqref{eq: con} can also be written as 
\begin{equation}\label{eq: subvliper}
\begin{aligned}
   V_L &< (\nu_{in}^{-1} v_l^3 l_\perp^{-1})^\frac{1}{2} \\
          &< 5.2~\text{km s}^{-1}  \Big(\frac{n_H}{0.2~\text{cm}^{-3}}\Big)^{-\frac{1}{2}} 
          \Big(\frac{v_l}{2.5~\text{km s}^{-1}}\Big)^\frac{3}{2} \\
         &~~~~\Big(\frac{l_\perp}{3.0\times10^{14}~\text{cm}}\Big)^{-\frac{1}{2}} ,
\end{aligned}
\end{equation}
{ with $\sigma(V_L) \sim 2.7~$km s$^{-1}$.}
The corresponding $M_A$ is 
\begin{equation}
\begin{aligned}
   M_A &= \frac{V_L}{V_{Ai}} < 0.13 \Big(\frac{n_e}{0.07~\text{cm}^{-3}}\Big)^{\frac{1}{2}} \Big(\frac{n_H}{0.2~\text{cm}^{-3}}\Big)^{-\frac{1}{2}} \\
        &  \Big(\frac{v_l}{2.5~\text{km s}^{-1}}\Big)^\frac{3}{2} 
         \Big(\frac{l_\perp}{3.0\times10^{14}~\text{cm}}\Big)^{-\frac{1}{2}} \Big(\frac{B}{ 5~\mu\text{G}}\Big)^{-1},
\end{aligned}
\end{equation}
{ with $\sigma(M_A)\sim 0.072$}.
Given such a small $M_A$ value, the implied $l_\text{tran} = L M_A^2$ 
{ is $\lesssim 6.1 \times 10^{12}~$cm.}
This means that 
the observed turbulence would be in the weak MHD turbulence 
regime with $l_\text{obs} > l_\text{tran}$. 
This is inconsistent with the observed Kolmogorov spectrum for strong MHD turbulence.  

\begin{figure*}[ht]
\centering
   \includegraphics[width=14cm]{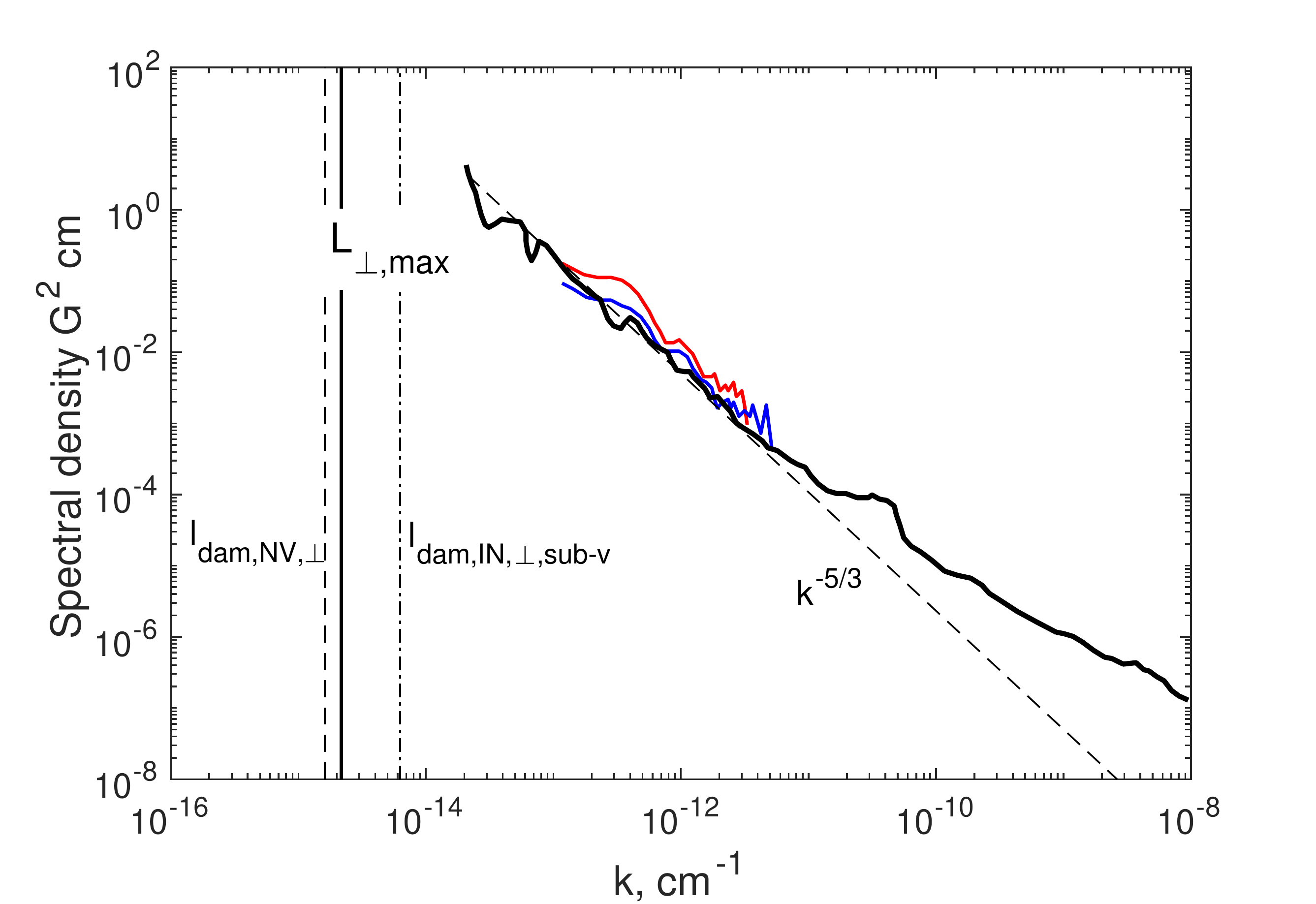}
\caption{
The data for the magnetic energy spectrum are taken from 
\citet{Lee20} (black) and
\citet{Burla18} (red and blue).
The dashed line indicates the Kolmogorov slope. 
The vertical dashed and dash-dotted lines correspond to the damping scales of interstellar turbulence 
($l_{\text{dam,NV},\perp}$, Eq. \eqref{eq:ismdamnv})
and interstellar magnetic fluctuations 
($l_{\text{dam,IN},\perp,\text{sub-v}}$, Eq. \eqref{eq:subvindam}).
The vertical solid line corresponds to 
the largest possible perpendicular injection scale 
($L_{\perp,\text{max}}$, Eq. \eqref{eq:anidlpi})
of the locally driven sub-Alfv\'{e}nic turbulence in the VLISM.}
\label{fig: spec}
\end{figure*}

{\it Case} (c).
{ In the above calculations we assume that the injection scale of turbulence is isotropic. 
In the presence of strong background magnetic fields, the driven sub-Alfv\'{e}nic turbulence is likely to have anisotropic $L$
(see e.g., \citealt{Pogo17}).
With a sufficiently small perpendicular component of injection scale $L_\perp$, the shear in the direction perpendicular to the magnetic field can cause significant distortions of magnetic field lines within the Alfv\'{e}n wave period. 
If the anisotropy is sufficiently large so that the nonlinear interaction between counterpropagating Alfv\'{e}n wave packets
is strong and thus the 
critical balance relation 
\citep{GS95}
is satisfied at $L$, the entire turbulent cascade would be in the strong MHD turbulence regime (see Fig. \ref{fig: tur}). 
In this case, we have 
\begin{equation}
    v_l = V_L \Big(\frac{l_\perp}{L_\perp}\Big)^\frac{1}{3},
\end{equation}
and 
$\Gamma_L = V_L / L_\perp$. The condition in Eq. \eqref{eq: con} leads to the constraint on the perpendicular injection scale,
\begin{equation}\label{eq:anidlpi}
\begin{aligned}
    L_\perp &< \Big(\nu_{in} v_l^{-1} l_\perp^\frac{1}{3}\Big)^{-\frac{3}{2}} \\
    &\approx 2.9\times10^{15}~\text{cm}
    \Big(\frac{n_H}{0.2~\text{cm}^{-3}}\Big)^{-\frac{3}{2}} 
  \Big(\frac{v_l}{2.5~\text{km s}^{-1}}\Big)^\frac{3}{2} \\
  &~~~~\Big(\frac{l_\perp}{3.0\times10^{14}~\text{cm}}\Big)^{-\frac{1}{2}}, 
\end{aligned}
\end{equation}
with the uncertainty $\sigma(L_\perp)\sim 1.5\times10^{15}~$cm,
and the same constraint on $V_L$ as in Eq. \eqref{eq: subvliper}.

Among the three cases, only {\it Case} (c), i.e., sub-Alfv\'{e}nic turbulence with anisotropic injection scale, provides the self-consistent result. 
In Fig. \ref{fig: spec}, we present  
$2\pi/L_{\perp,\text{max}}$ together with the observationally measured magnetic energy spectrum taken from 
\citet{Lee20} for the period from 2012 August to 2019 December
and
\citet{Burla18} for intervals 
2013.3593-2014.6373 and 2015.3987-2016.6759.
Here $L_{\perp,\text{max}}$ (Eq. \eqref{eq:anidlpi}) is the largest possible perpendicular injection scale of the LMHD turbulence in the VLISM.
We find that $L_{\perp,\text{max}}$ is close to 
$l_{\text{dam,NV},\perp}$ of the interstellar turbulence given by Eq. \eqref{eq:ismdamnv}.}

\section{Discussion}

\subsection{New regime of MHD turbulence}
\label{ssec: newmhd}

In the case when the NV damping dominates over the IN damping (see Section \ref{ssec: nvin}), 
the turbulent motions are damped at the neutral viscous scale, but the magnetic fluctuations can exist on smaller scales. 
There is a new regime of MHD turbulence in the sub-viscous range \citep{CLV_newregime,LVC04,XL16},
where the kinetic energy spectrum is steep with the spectral index $-4$ and the magnetic energy spectrum is flat with the spectral index $-1$.
We note that such a flat magnetic energy spectrum corresponds to scale-independent magnetic fluctuations.
The sub-viscous magnetic fluctuations are caused by the shear of viscous-scale turbulent eddies in the direction perpendicular to the magnetic field. 
As the IN damping suppresses magnetic fluctuations, the damping scale of the sub-viscous magnetic fluctuations is determined by the balance between the eddy-turnover rate at $l_{\text{dam,NV},\perp}$ and the IN damping rate,
\begin{equation}\label{eq: blansub}
 \frac{v_\nu}{l_{\text{dam,NV},\perp}} = \omega_\text{d,IN},
\end{equation}
{ where $v_\nu = V_{L,\text{ISM}}(l_{\text{dam,NV},\perp}/L_\text{ISM})^{1/3}$ is the turbulent velocity at $l_{\text{dam,NV},\perp}$.}
The IN damping rate in this case is
\begin{equation}\label{eq:indasubv}
\omega_\text{d,IN} = \frac{\xi_n k_\perp^2 (\delta V_A)^2}{2 \nu_{ni}},
\end{equation}
where 
$\delta V_A = \delta B /\sqrt{4\pi\rho}$, and 
$k_\perp^2 (\delta V_A)^2$ 
is the magnetic force per unit mass per unit displacement 
corresponding to the sub-viscous magnetic fluctuation $\delta B$ perpendicular to the magnetic field. 
Eq. \eqref{eq:indasubv} is different from the expression for IN damping rate of Alfv\'{e}n waves in
\citet{KulA92}, as
over sub-viscous scales, there are no Alfv\'{e}n wave motions of magnetic fields. 
By assuming that the Alfv\'{e}nic component 
{ dominates the turbulent motion},
we approximately have 
$\delta V_A \approx v_\nu$.
Then we obtain from Eq. \eqref{eq: blansub}
\begin{equation}\label{eq:subvindam}
\begin{aligned}
 l_{\text{dam,IN},\perp,\text{sub-v}}&=\frac{\xi_n}{2}  \nu_{ni}^{-\frac{1}{2}}\nu_n^\frac{1}{2} \\
 &\approx 1.0\times10^{15}~\text{cm} \Big(\frac{T}{6300~\text{K}}\Big)^\frac{1}{4} \Big(\frac{n_e}{0.07 ~\text{cm}^{-3}}\Big)^{-\frac{1}{2}}\\
 &~~~~\Big(\frac{n_H}{0.2 ~\text{cm}^{-3}}\Big)^{-\frac{1}{2}},
\end{aligned}
\end{equation}
{ with $\sigma(l_{\text{dam,IN},\perp,\text{sub-v}})\sim 8.5\times10^{13}~$cm,}
as the IN damping scale of sub-viscous magnetic fluctuations
(see Fig. \ref{fig: spec}). 
We see that $l_{\text{dam,IN},\perp,\text{sub-v}}$ is slightly smaller than 
$l_{\text{dam,NV},\perp}$
in Eq. \eqref{eq:ismdamnv}.
So the cutoff scale of magnetic fluctuations is close to the cutoff scale
of the kinetic energy spectrum of the interstellar turbulence. 


{ 
For the LMHD turbulence driven in the VLISM, the new regime of MHD turbulence is also expected below the (effective) ion viscous scale. 
As pointed out in 
\citet{Frat21},
the small-scale spectral flattening with the spectral index $-1$ 
(see also Fig. \ref{fig: spec})
may be accounted for by the sub-viscous magnetic fluctuations in the new regime of LMHD turbulence 
\citep{CLV_newregime,XL16}.}

\subsection{Compressibility of the LMHD turbulence}

In our calculations,
we assume that the Alfv\'{e}nic component carries most of the turbulent energy in compressible MHD turbulence. This is supported by compressible MHD turbulence simulations 
\citep{CL02_PRL,Hu22},
{ as well as the higher power of perpendicular magnetic fluctuations compared to that of parallel magnetic fluctuations found in 
\citet{Lee20}.
We consider strong MHD turbulence with 
strong nonlinear interactions between oppositely directed Alfv\'{e}n wave packets and 
balanced cascade 
\citep{GS95}
based on the Kolmogorov magnetic energy spectrum reported in 
e.g., \citet{Burla18,Lee20}.
But we also note the existence of 
large-scale compressive component of magnetic
fluctuations with the spectral index close to $-2$
\citep{Frat21},
which probably reflects the discontinuities in magnetic field distribution associated with shock/compression waves.
As the conversion from Alfv\'{e}n modes to compressive modes of MHD turbulence is inefficient
\citep{CL02_PRL},
the compressive component is likely to originate from the turbulence injection.
By combining the in-situ data of magnetic and electron density fluctuations,
\citet{Lee20}
found that the observations cannot be explained by the linear magnetohydrodynamic modes alone.
In addition, within the limited range of measured frequencies/scales, it is difficult to distinguish between the spectral indices
$-5/3$ and $-2$ in observations, and thus the 
possibility of the existence of weak turbulence cannot be completely excluded.}

{ Due to the damping effects in the presence of neutrals, we find that the measured turbulence in the VLISM is likely to be entirely of heliospheric origin.
Moreover,
the turbulent energy transfer rate of the LMHD turbulence is 
$\epsilon_\text{LMHD}\sim \rho_i V_L^3 /L_\perp \approx 5.7\times 10^{-24}~$erg cm$^{-3}$ s$^{-1}$, which is much larger than 
$\epsilon_\text{ISM}\sim \rho V_{L,\text{ISM}}^3/L_\text{ISM} \approx 1.4\times10^{-26}~$erg cm$^{-3}$ s$^{-1}$
of the interstellar turbulence.
This is consistent with the finding in 
\citet{Lee20,Ocke21}
that the Kolmogorov electron density spectrum measured by {\it Voyager 1} has a significantly higher intensity than that
measured in the interstellar warm ionized medium 
\citep{Armstrong95,CheL10}.
The quasiperiodic structures
seen in the magnetic field fluctuations also indicate the 
heliospheric forcing of the measured turbulence in the VLISM
\citep{Frat21}.
Both solar rotation and
solar cycle may play an important role in driving the heliospheric
turbulence 
\citep{Zank19}.}

With large observational uncertainties, 
the transition from the magnetic fluctuations parallel to the mean magnetic field in an earlier interval to those transverse to the mean magnetic field in later intervals was found from {\it Voyager 1} and {\it 2} measurements 
\citep{Burla18,Zhao20,Bur22}. 
Magnetic fluctuations parallel to the local magnetic field arising from slow and fast modes in compressible MHD turbulence and pseudo-Alfv\'{e}nic modes in incompressible MHD turbulence are important for the mirror diffusion of particles
\citep{XL20,LX21}.
\cite{Zank19} has shown that the fluctuations measured by 
{\it Voyager 1} and {\it 2} can be explained by a transmission of fast modes from inner
heliosheath with a conversion to incompressible Alfv\'{e}n waves. They further  
assumed that the the outer scale of this heliosphere-originated turbulence is around 
120 au, and the interstellar turbulence will continue down to scales smaller than
those measured by {\it Voyager}. 
Our work, by considering  the ion-neutral interactions in the LISM
region, suggests that the interstellar turbulence should have a cutoff 
around $261$ au, 
and a new heliosphere-related turbulence (such as the scenario 
described by \cite{Zank19}) with an injection scale smaller than
{ $\sim 2.9\times 10^{15}$ cm ($\approx 194$ au)} is needed to explain the {\it Voyager} measurements.


\section{Conclusions}

The damping effects in the partially ionized LISM determine the range of length scales for the existence of interstellar MHD turbulence and the LMHD turbulence. 
Due to the high temperature and moderate ionization fraction in the LISM, 
we find that the dominant damping mechanism of the interstellar MHD turbulence is the NV damping. 
{ The NV damping scale of the interstellar turbulence is about $261$ au.}
Below the NV damping scale of turbulent cascade, 
the new regime of MHD turbulence with constant magnetic fluctuations
\citep{LVC04}
is expected to rise. 
We find that the sub-viscous magnetic fluctuations are cut off due to the IN damping at a scale slightly smaller than the NV damping scale.

For the LMHD turbulence, 
when the injection occurs in the regime with ions decoupled from neutrals, 
the turbulent cascade can not be cut off by the damping related to partial ionization.
{ Given the turbulent velocity at the largest observed length scale} inferred from in-situ measurements, 
after applying 
the ion-neutral decoupling condition, 
we find that the LMHD turbulence in the VLISM is sub-Alfv\'{e}nic
with the injected turbulent energy smaller than the magnetic energy. 
{ With the trajectory of {\it Voyager 1} approximately perpendicular to the background magnetic field, 
by assuming anisotropic injection scale of the LMHD turbulence, 
we further find the upper limit of the perpendicular injection scale $L_{\perp,\text{max}} \approx 2.9\times10^{15}$ cm $\approx 194$ au, 
which is close to the NV damping scale of the interstellar turbulence.} 
Our estimated largest outer scale of LMHD turbulence is comparable to the 
extent of the heliosphere in the upstream direction 
\citep{Pogo17}
and one order of magnitude smaller than the estimate given in 
\citet{Burla18,Lee20}
by extrapolating the power-law slope to 
{ the equipartition between the turbulent fluctuation and the average magnetic field strength}.
Note that 
other considerations such as mode conversion \citep{Zank19} and IBEX modeling
\citep{Zirn20} could 
further limit the injection scale down to tens of au. 

\acknowledgments
{ We thank the referees for very detailed and helpful comments, which improved the manuscript perceptibly.}
S.X. acknowledges 
inspiring discussions with Alex Lazarian, Ethan Vishniac, and 
the support for 
this work provided by NASA through the NASA Hubble Fellowship grant \# HST-HF2-51473.001-A awarded by the Space Telescope Science Institute, which is operated by the Association of Universities for Research in Astronomy, Incorporated, under NASA contract NAS5-26555. 
H.L. acknowledges useful discussions with 
Fan Guo, Eric Zirnstein, and 
the support by LANL LDRD program. 
\software{MATLAB \citep{MATLAB:2021}}

\appendix

\section{Anisotropy of strong MHD turbulence}
\label{app:ani}

The theoretically established scale-dependent anisotropy of trans-Alfv\'{e}nic turbulence
\citep{GS95} and 
sub- and super-Alfv\'{e}nic turbulence 
\citep{LV99}
has been tested by MHD turbulence simulations 
\citep{CL02_PRL,CL03}
and supported by 
spacecraft measurements in the solar wind
\citep{Horb08,Luo10,For11}.
Here we briefly review the anisotropy of strong Alfv\'{e}nic turbulence. 

The cascading rate of Alfv\'{e}nic turbulence in strong MHD turbulence regime is 
\begin{equation}\label{eq:casrsa}
    \tau_\text{cas}^{-1} = v_l l_\perp^{-1} = V_\text{st} L_\text{st}^{-\frac{1}{3}}l_\perp^{-\frac{2}{3}}, 
\end{equation}
where $V_\text{st}$ is the turbulent velocity at the outer scale $L_\text{st}$ of strong MHD turbulence. More specifically, there is 
\begin{equation}
    V_\text{st} = V_A, 
    L_\text{st} = l_A= L M_A^{-3}
\end{equation}
for super-Alfv\'{e}nic turbulence with $M_A>1$, 
and 
\begin{equation}
    V_\text{st} = V_L M_A, 
    L_\text{st} =l_\text{tran}= L M_A^2
\end{equation}
for sub-Alfv\'{e}nic turbulence with $M_A<1$,.
We note that $V_A$ should be replaced by $V_{Ai}$ when the turbulence is injected in the decoupled regime. 

By combining the expression of $\tau_\text{cas}^{-1}$ with the critical balance relation 
\begin{equation}
    \tau_\text{cas}^{-1} = \frac{V_A}{l_\|}, 
\end{equation}
we can obtain the anisotropic scaling relation of strong Alfv\'{e}nic turbulence, 
\begin{equation}
    l_\| = \frac{V_A}{V_\text{st}} L_\text{st}^\frac{1}{3}l_\perp^\frac{2}{3}.
\end{equation}
It shows that smaller-scale turbulent eddies are more elongated
along the local magnetic field.

\section{Damping scales of MHD turbulence cascade in a partially ionized medium}
\label{app:dam}

Under the consideration of both IN and NV damping effects in a partially ionized medium, 
\cite{XLY14,Xuc16,XLp17}
derived the general expression of the damping rate of Alfv\'{e}nic turbulence. Here we briefly review its approximate form in different coupling regimes. 

In the weakly coupled regime, there is only the IN damping, with the damping rate 
\begin{equation}
    \omega_d = \frac{\nu_{in}}{2}.
\end{equation}
For the MHD turbulence injected in the decoupled regime, as the cascading rate is always larger than $\nu_{in}$ and thus larger than $\omega_d$, the cascade is not cut off by the damping in a partially ionized medium. 

In the strongly coupled regime, the damping rate can be approximately written as 
\begin{equation}\label{eq: damstbo}
    \omega_d = \frac{\xi_n}{2} \Big(k^2 \nu_n + \frac{k_\perp^2 (\delta V_A)^2}{\nu_{ni}}\Big)
    =\frac{\xi_n}{2} \Big(k^2 \nu_n + \frac{k_\|^2 V_A^2}{\nu_{ni}}\Big),
\end{equation}
where we assume that the magnetic fluctuations are mainly induced by Alfv\'{e}nic turbulence and apply the critical balance relation for strong MHD turbulence, 
$k_\perp \delta V_A = k_\perp v_k = k_\| V_A$.
Here $v_k$ is the turbulent velocity at wavenumber $k$.
The first and second terms of $\omega_d$ correspond to NV and IN damping, respectively. 

For the MHD turbulence injected in the strongly coupled regime, when the damping rate exceeds the cascading rate, MHD turbulence cascade is damped. 
We consider that the damping scale is in the strong MHD turbulence regime. 
By comparing Eq. \eqref{eq: damstbo} with Eq. \eqref{eq:casrsa},
we find the damping scale 
\begin{equation}
    l_{\text{dam,NV},\perp} = \Big(\frac{\xi_n}{2}\Big)^\frac{3}{4} \nu_n^\frac{3}{4} L_\text{st}^\frac{1}{4} V_\text{st}^{-\frac{3}{4}}
\end{equation}
when the NV damping dominates over the IN damping, 
where we assume $k\sim k_\perp$,
and 
\begin{equation}
    l_{\text{dam,IN},\perp} = \Big(\frac{2\nu_{ni}}{\xi_n}\Big)^{-\frac{3}{2}} L_\text{st}^{-\frac{1}{2}} V_\text{st}^\frac{3}{2}
\end{equation}
in the opposite case. 
If the MHD turbulence is damped due to neutral viscosity in the strongly coupled regime, there is the so-called new regime of MHD turbulence on scales below the NV damping scale 
\citep{LVC04}.
If the damping of MHD turbulence is caused by neutral-ion decoupling, hydrodynamic turbulent cascade can happen in neutrals that are decoupled from ions on scales smaller than the IN damping scale
\citep{XLY14,Buam15}.

\bibliographystyle{aasjournal}
\bibliography{xu}

\end{CJK*}
\end{document}